\documentclass[conference]{IEEEtran}
\IEEEoverridecommandlockouts
\usepackage{cite}
\usepackage{amsmath,amssymb,amsfonts}
\usepackage{algorithmic}
\usepackage{graphicx}
\usepackage{textcomp}
\usepackage{xcolor}
\def\BibTeX{{\rm B\kern-.05em{\sc i\kern-.025em b}\kern-.08em
    T\kern-.1667em\lower.7ex\hbox{E}\kern-.125emX}}
\begin{document}

\title{A Smart Background Scheduler for Storage Systems\\
}

\author{\IEEEauthorblockN{Maher Kachmar}
\IEEEauthorblockA{\textit{Northeastern University} \\
\textit{Department of Electrical and Computer Engineering}\\
Boston, MA \\
kachmar.ma@northeastern.edu}
\and
\IEEEauthorblockN{David Kaeli}
\IEEEauthorblockA{\textit{Northeastern University} \\
\textit{Department of Electrical and Computer Engineering}\\
Boston, MA \\
kaeli@ece.neu.edu}
}

\maketitle

\begin{abstract}
In today's enterprise storage systems, supported data services such as snapshot delete or drive rebuild can cause tremendous performance interference if executed inline along with heavy foreground IO, often leading to missing SLOs (Service Level Objectives). Typical storage system applications such as web or VDI (Virtual Desktop Infrastructure) follow a repetitive high/low workload pattern that can be learned and forecasted. We propose a priority-based background scheduler that learns this repetitive pattern and allows storage systems to maintain peak performance and in turn meet service level objectives (SLOs) while supporting a number of data services. When foreground IO demand intensifies, system resources are dedicated to service foreground IO requests and any background processing that can be deferred are recorded to be processed in future idle cycles as long as forecast shows that storage pool has remaining capacity. The smart background scheduler adopts a resource partitioning model that allows both foreground and background IO to execute together as long as foreground IOs are not impacted where the scheduler harness any free cycle to clear background debt. Using traces from VDI application, we show how our technique surpasses a method that statically limit the deferred background debt and improve SLO violations from 54.6\% when using a fixed background debt watermark to merely a 6.2\% if dynamically set by our smart background scheduler.

\end{abstract}


\section{Introduction}
Storage servers and the cloud providers are today more than ever under marketing pressure to meet their service level objectives (SLOs) and maintain peak\footnote{Storage Server Peak Performance is measured under favorable conditions where no data services are running}  performance while supporting ever-exploding large datasets. Due to advances in hardware in general and storage processors in particular, hundreds of cores can be packed in multiple CPU sockets and hundreds of gigabytes of memory are typical of nowadays storage processors, which is a building block for storage clusters \cite{mit,dell}. It is expected that storage providers utilize these resources to scale to larger datasets, exceed a higher performance bar, all while supporting services such as data protection, data efficiencies, and data analytics. For instance, if volume snapshot retention time happen to expire during a burst of foreground IO or a request to unmap a large address space was issued by an application or a disk needs to be rebuilt during a ‘busy’ period, this can stress the storage system and lead to QoS violations. A recent study \cite{Qiao2019} reports that drive rebuild has 78.5\% impact on the performance of ZFS file system.

Furthermore, storage vendors are adopting log structured based design that improves write workloads through coalescing multiple writes into a larger data block while deferring garbage collection of old overwritten data to a background process \cite{wiki01}. This log structured design often utilizes a fast but expensive media tier that requires a background process to continuously flush newly written data to a slower media tier. This background process is considerably ‘heavy’ compared to foreground IO especially if it has to provide core data efficiency services such as inline compression and deduplication.

Moreover, during load bursts, the foreground I/O stack can be further lightened by deferring any data/metadata, that needs be prefetched and/or modified in order to complete an I/O, to another background process so that more IOPS can be squeezed during said bursts. 

Storage application background processing include the following types of background ops but they are not limited to this list:

\begin{itemize}
\item Snap shot delete, volume delete , and unmaps.
\item Deep compression and offline deduplication. 
\item Garbage collection and defragmentation.
\item Drive / stripe rebuild.
\item Prefetching and relocation of data.
\item Data and Metadata flush.
\item Replication sync.
\item Integrity checks.
\item Data Analytics.
\end{itemize}
Except the last 3 types of background ops which improves reliability and serviceability, the only drawback of not instantaneously catering for these background ops is the potential to tie free pool storage. However, tying up too much deferred debt can lead to pool running out of space which is another ballpark hard SLA violation, DU/DL (Data Unavailable/Data Loss). Such a requirement justifies coming up with a smart background scheduler that is powered by forecasting models, statistical or machine learning based, to lively profile and characterize workloads on a per deployment basis in order to strike a perfect balance between servicing user IO requests and meeting the required data services.

Our proposed smart background scheduler adopts a resource partitioning model that allows both foreground and background IO to execute together as long as foreground IOs are not impacted. The scheduler is backed by a forecasting component so to harness any free cycle to clear the background debt and keep background work minimal during heavy IO. Using traces from VDI (Virtual Desktop Infrastructure) application, we show how our technique surpasses a method that statically limit the deferred background debt and improve SLO violations from 54.6\% when using a fixed background debt watermark to merely a 6.2\% if dynamically set by our smart background scheduler.
 
After this introduction, the paper is organized in the following order. Section 2 discusses related work. In Section 3, we present the experimental setup. Section 4 presents motivation and a workload characterization of multiple popular applications hosted on a typical enterprise storage system. In Section 5, we present the forecasting and the background scheduler models. Section 6 presents trace driven simulations that provides a comparison of multiple forecasting algorithms and demonstrates the effectiveness of our background scheduler. Finally, we conclude and list future work in Section 7.
\\

\section{Related Work}
There is a rich body of work already established on workload characterization and predictions. They are typically used for cache warmup and prefetching \cite{Xue2014}. In some cases, these workload characterizations have been used to direct the scheduling of maintenance or background work \cite{Riska2006, Yan2012, Eggert2005}, for capacity planning \cite{Stokely2012}, or for improving the performance of busty workloads \cite{Mi2012}. Taylor \cite{Taylor2017}, et. al., lays down a time-series forecasting method that can be used for most datasets with limited data science expertise. Their method decompose time-series to 3 components: trend, seasonality, and holiday and address each component separately. Their “prophet’ forecasting model is able to elegantly capture short and long term seasonality as a curve-fitting exercise, all without over-fitting training data. Alshawabkeh \cite{Alshawabkeh2012}, et. al., uses Markov Chain Correlation of the spacial and temporal characteristics of storage blocks to intelligently place blocks on a tiered storage systems. Their method isolate a group of devices into a cluster that behave similarly. A Markov Chain Cluster transition between High, Medium, and Low arrival intensities which in turn guide the placement of these devices on their ideal tier. Ravandi \cite{Ravandi2017, Ravandi2017a}, et. al., models storage providers on cloud as a black box where they assess vendors’ level of Quality of Service (QoS) through monitoring or synthetic test tool. In turn, this information is fed to a ML based algorithm where they classify the level of QoS met in order to enhance SLA/SLO violations and balance it with storage capacity requested. Xue \cite{Xue2014, Xue2016}, et. al, explains how to use machine learning techniques such as neural networks to predict user workload intensities in order to schedule data analytics work during idle time but also warmup caches before the next cycle of user workload. Stokely \cite{Stokely2012}, et. al, used ensembles of time series models to project disk utilization trends in a cloud setting. Zhang \cite{Zhang2006}, et. al., study the IO characteristics of storage systems to ensure that reliability background jobs do not impact foreground IO. Their analytical model incorporates characteristics such as burstiness, arrival dependence, and system utilization and use it to find optimal intersection points where impact of background jobs on foreground IO is reduced.

However, we depart from previous work in several aspects that allow us to provide a holistic and more realistic solution for the storage industry. Prior approaches assumed that background ops cannot be executed along with foreground IO or the need to warmup the cache after executing each workload. With recent advances in hardware where a datacenter-on-a-chip is no longer a science fiction \cite{mit}, this paradigm is no longer applicable.  Isolating execution domains between foreground and background allows these different workloads to co-execute in their perspective domains. Moreover, prior approaches lack key information needed for the scheduler to perform as an ideal storage cluster that has both performance and capacity guarantees. It is not feasible to delay background processing if the pool is running out of blocks or when reliability becomes compromised when not meeting drive rebuild time. To provide a framework for the storage industry where performance and capacity are tightly coupled, we feed this knowledge directly to the priority based scheduler where we: 
\begin{itemize}
\item Quantify the cost of the deferred background work and the time to clear it. 
\item Track free pool space and deferred debt tied space in order to estimate time before running out of resources.
\item Grade current incoming load and harvest any free cycle to execute background ops using isolated domains to reduce any effect on foreground I/O.
\end{itemize}
For example, deleting hundreds of snaps is typically CPU intensive, but also, deferring large volume deletes or deferring offline deduplication can tie tremendous storage capacity risking the user application to run out of space. Adding these key information to the model helps provide a more realistic priority based scheduler for the storage server that balance both foreground and background processing.

\section{Experimental Setup}
To guide the development our smart scheduler, we run synthetic workloads on a live storage server.  Our testbed consists of a server running Centos 7 with 2 NUMA CPU sockets, 6 cores per socket, 24 GB of memory and 2 SAS2 drives.  The SAS2 drive have a capacity of 300GB and can support 10,500 RPM. We use the FIO tool version 3.7 to produce synthetic block-based workload on the system~\cite{fio}. FIO has a capability to create various workload streams (random, mixed, etc.) with varying compression and deduplication ratios. This tool is used to produce workload traces used to evaluate the scheduler.

Various storage traces of real enterprise applications exist. SNIA~\cite{snia} provide VDI traces (discussed in later section), as well several other storage traces. Wikipedia traces also exist for public research use~\cite{Urdaneta2009}. These traces are used to compare the forecasting models, which are used in turn to evaluate our smart background scheduler.

After collecting traces, we designed a trace-driven simulator that has the following features:

\begin{itemize}
\item Ability to process I/O traces.
\item Ability to read and apply background scheduling policies (snap policies, efficiency policies, etc.).
\item Model background processing costs for various background workload types.
\item Modular design with ability to attach to various background scheduler models (be able to attach to any Forecasting or Scheduler model).
\item Generate performance metrics and visualize results.
\end {itemize}

Our goal is to compare the performance of the system with and without background data services supported. Ideally, the background data services should have little to no impact on the user during IO busy phases. The proposed solution should surpasses any method that statically limit the deferred background debt.

\section{Trace Generation}

\begin{figure}[h!]
  \includegraphics[height=4cm]{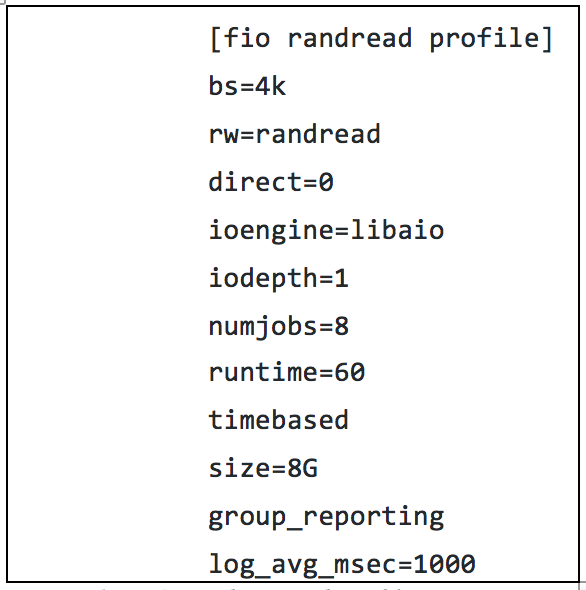}
  \caption{FIO random read profile.}
  \label{fig:fioprofile}
\end{figure}

\begin{figure}
  \includegraphics[width=0.5\textwidth]{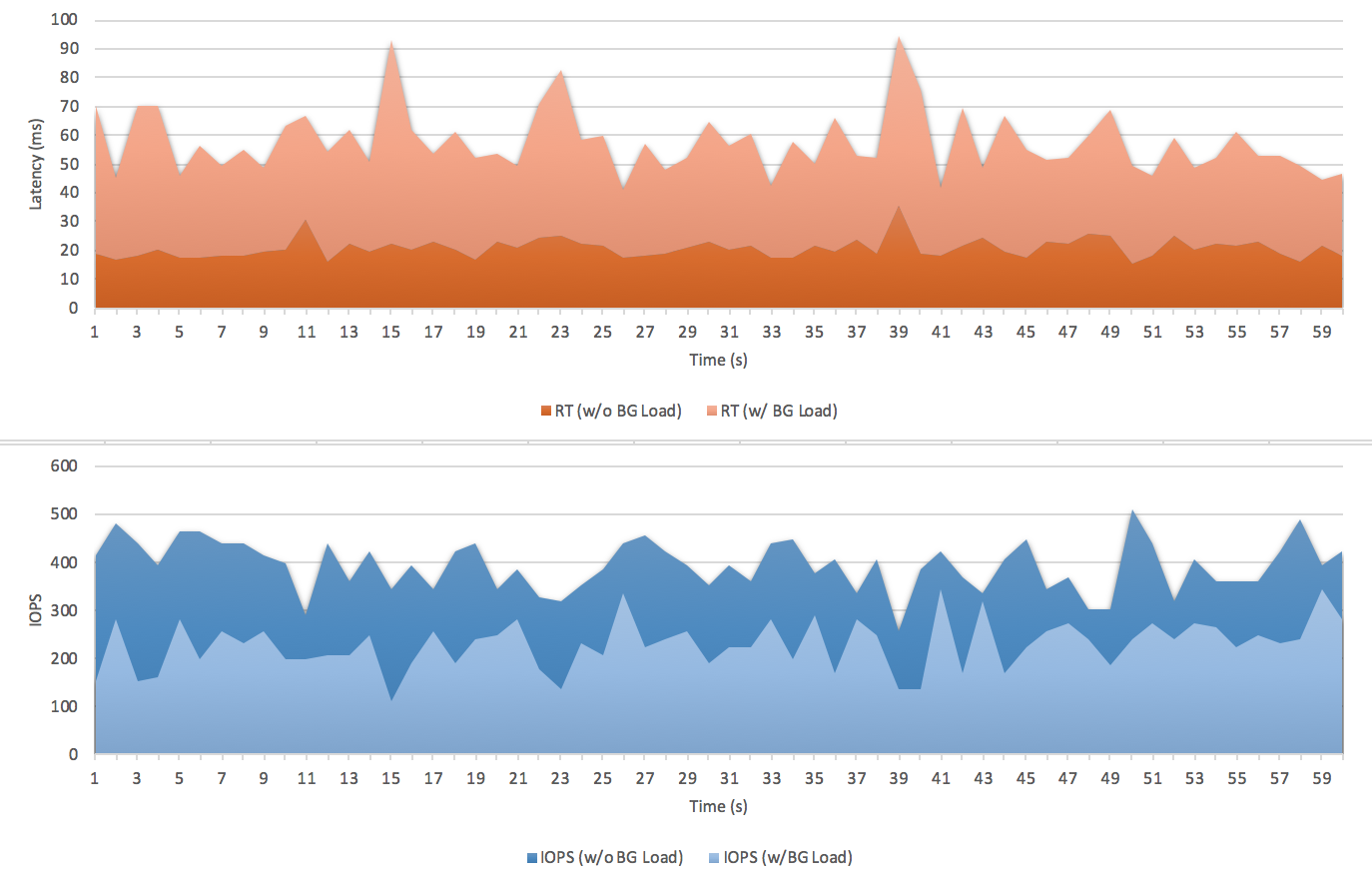}
  \caption{Impact of background load on the performance of the XFS filesystem.}
  \label{fig:BGLoad}
\end{figure}

To better understand the potential performance degradation introduced by background processing, for the Linux filesystem (XFS) on the testbed we ran FIO \cite{fio} to generate a random read workload using 8 jobs (threads).   This workload nearly saturates the drives ($\geq{90\%}$). We consider here two scenarios. The first test is run without any background load. Then in the second test introduce some background load. We represented the background load in this experiment by concurrently running deletes of 20 large files. Figure~\ref{fig:fioprofile} shows the FIO profile used in this experiment. Figure~\ref{fig:BGLoad} reports the IOPS and resulting response time on XFS file system. 

Although the delete of these 20 files may have only involved metadata updates in this filesystem implementation, this background traffic was still enough to produce a measurable impact. We found that IO latency was doubled and the number of IOPS were cut in half. This impact grows significantly in the case when the background process is a snap delete of a volume that shares data blocks between the deleted snap and primary copy.

Today, a typical storage array of same family and comparable feature set is sized and priced based on two criteria:
\begin{enumerate}
\item Performance - the maximum IOPS achieved and the ability to handle bursty traffic due to virtual machine boots or shutdown storms.
\item Capacity - the number of objects, physical space, and efficiency achieved through thin provisioning, compression, and deduplication.
\end{enumerate}
Thus, the goal of our smart background scheduler is to tradeoff capacity in order to gain performance, as long as there is some excess performance available during periods.
\begin{figure}[h!]
  \includegraphics[width=0.5\textwidth]{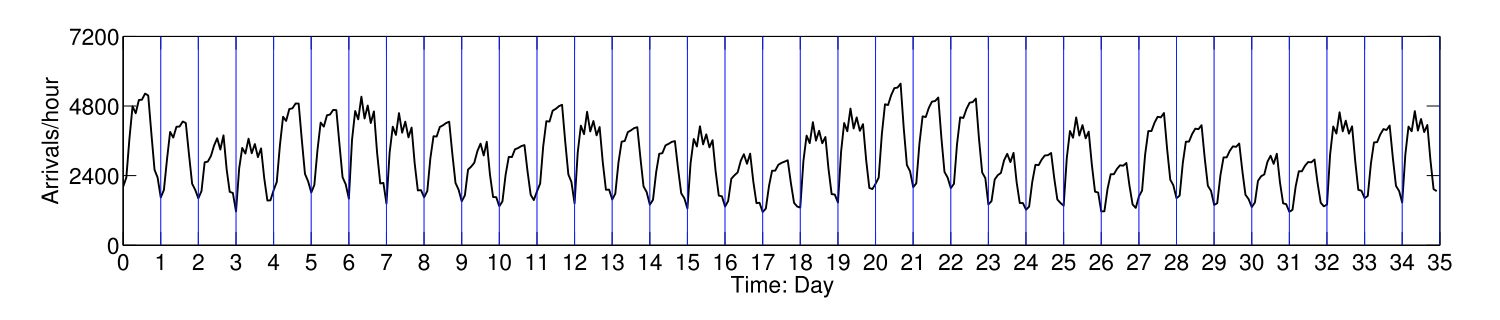}
  \caption{Typical storage system IO arrival rate over a 35 day period \cite{Xue2014}.}
  \label{fig:intensityarrival}
\end{figure}
\begin{figure}[h!]
  \includegraphics[width=0.5\textwidth]{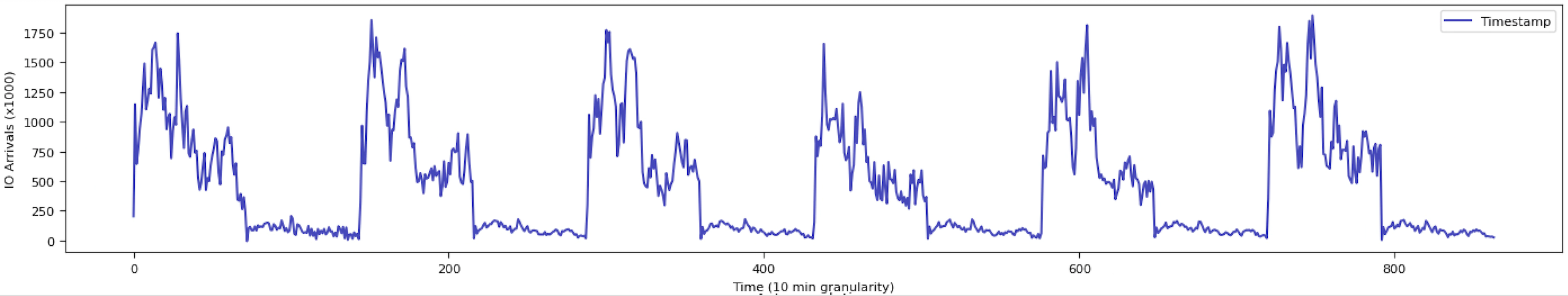}
  \includegraphics[width=0.5\textwidth]{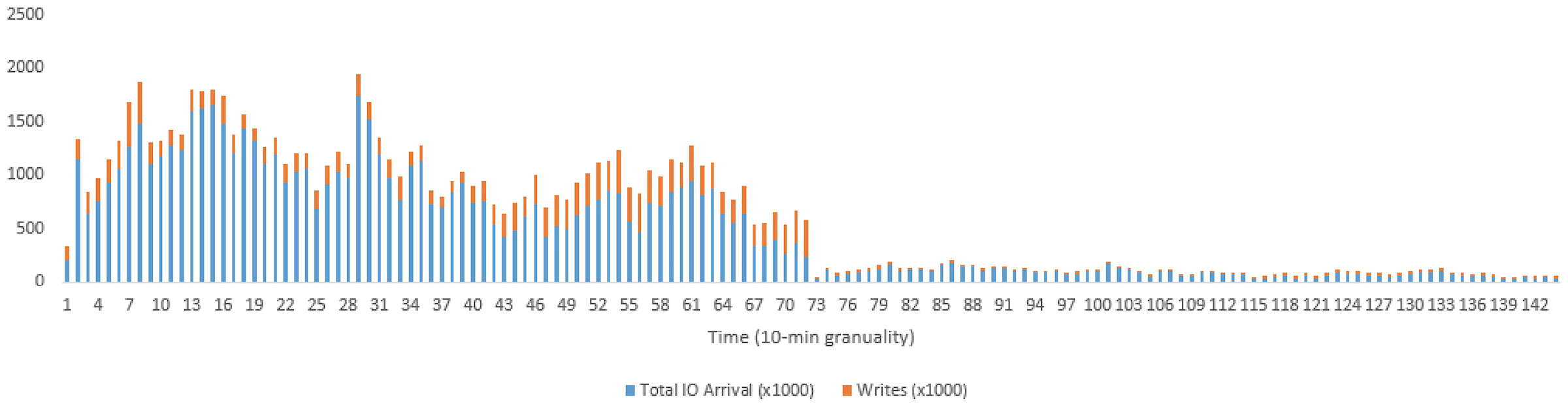}
  \caption{User request arrival intensity (Incoming load) of VDI workload on 3 LUNs over 6 days (starting 2/24/2016 @ 7am) and a zoomed-in view of the first 24-hour period.}
  \label{fig:intensityarrival2}
\end{figure}

\begin{figure}[h!]
  \includegraphics[width=0.41\textwidth]{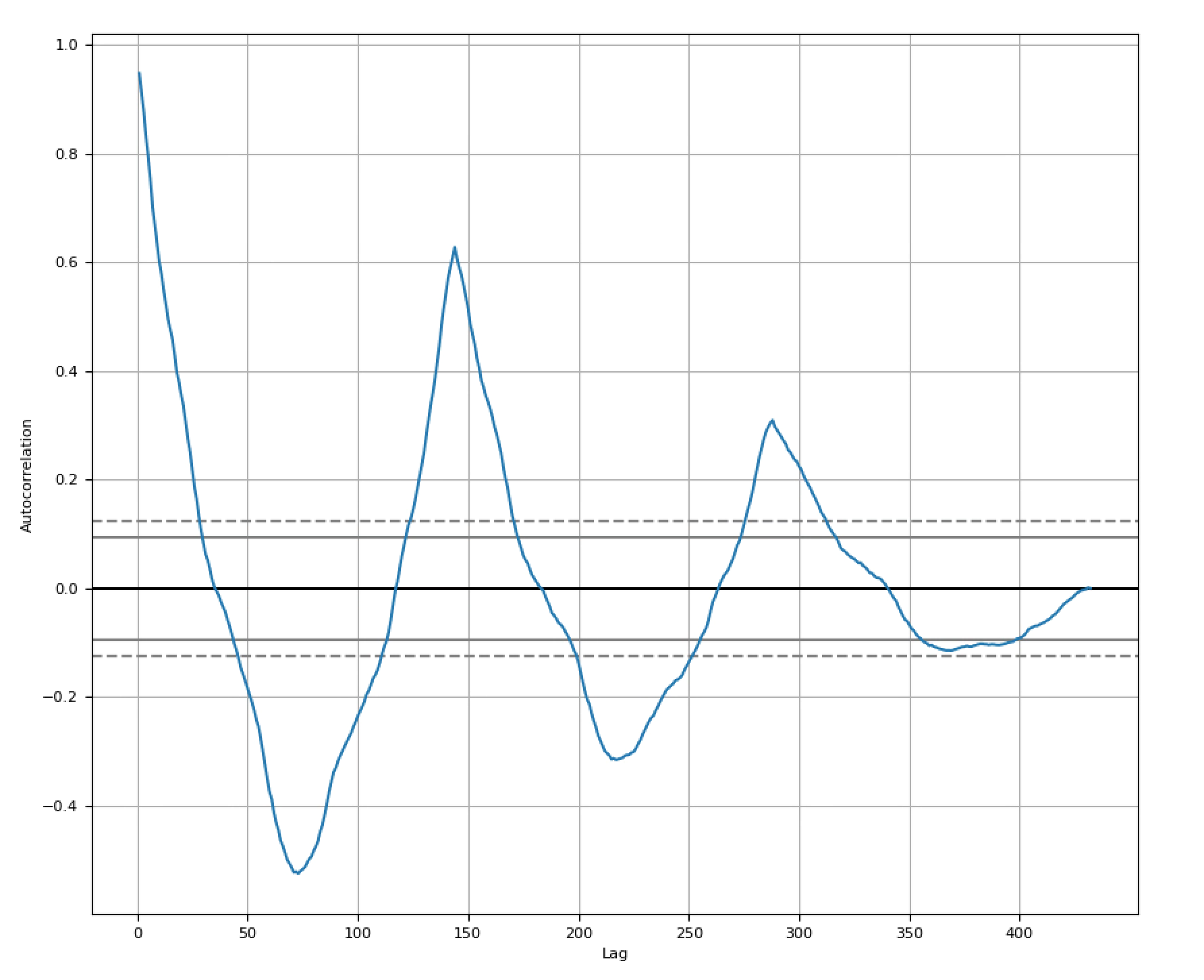}
  \caption{VDI workload autocorrelation at different lags.}
  \label{fig:Autocorrelation}
\end{figure}

\begin{figure}[h!]
  \includegraphics[width=0.41\textwidth]{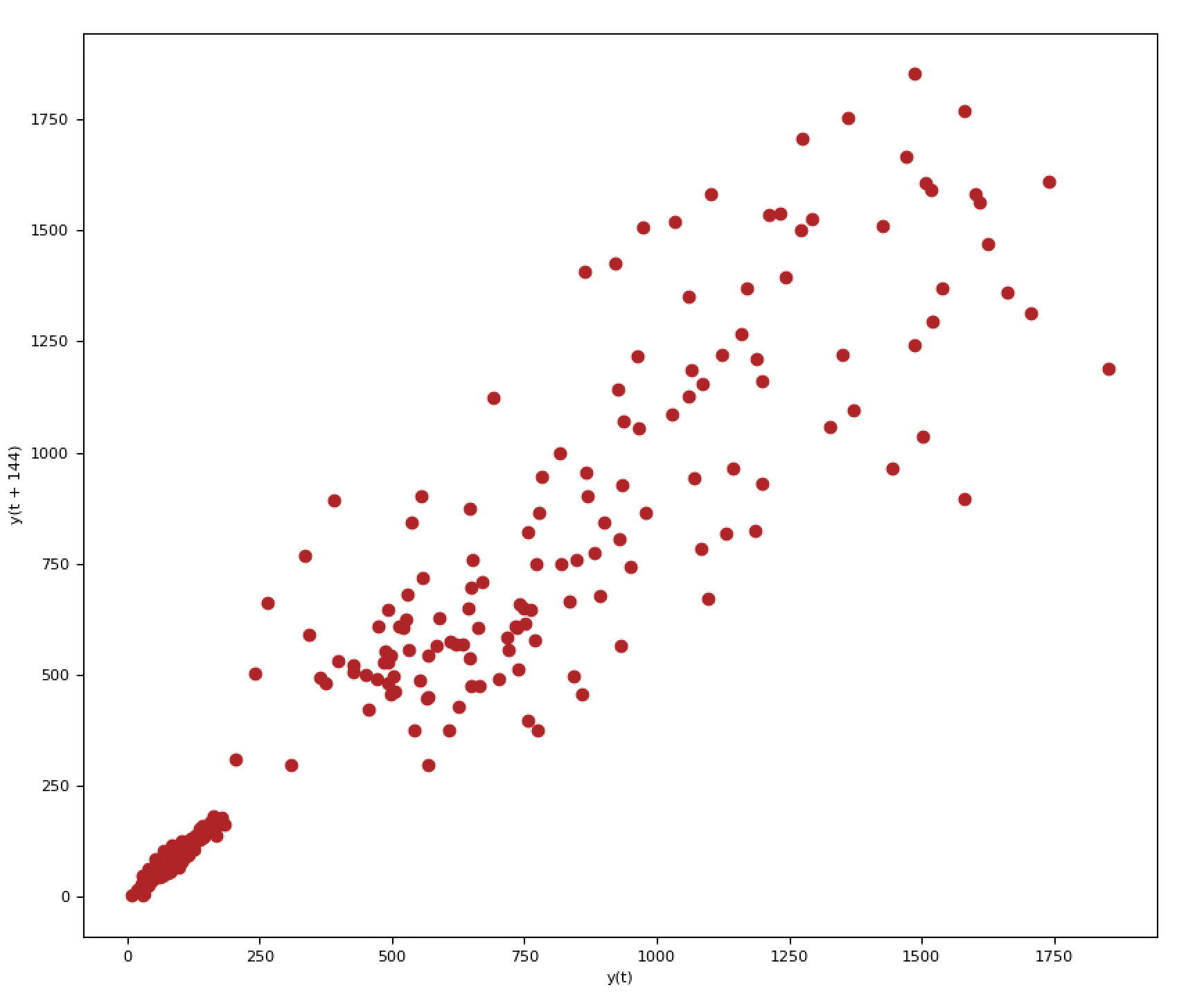}
  \caption{24-hour lag spread of VDI workload.}
  \label{fig:lagspread}
\end{figure}

Many storage workloads tend to be bursty in nature, follow repeatable patterns in time (day/night or weekday/weekend), and are tightly coupled to the number of time zones in which storage clusters are deployed. Figure~\ref{fig:intensityarrival} shows the IO arrival rate for a production storage system \cite{Xue2014}. There are clear patterns present, based on day/night and weekday/weekend intervals. Figure~\ref{fig:intensityarrival2} shows the input IO intensity of our Virtual Desktop Infrastructure (VDI) workload run on 3 LUNs over a 6 day period.  We include a snapshot  of the first 24-hour period . The traces can be obtained from SNIA \cite{snia}. Even at a fine granularity (hourly), there clearly a distinct pattern that repeats. Figure~\ref{fig:Autocorrelation} plots the autocorrelation of incoming intensity at different lags, where the highest at is at point 144 (a one-day lag, plotted at 10 minute intervals). Figure~\ref{fig:lagspread} shows the 24-hour lag spread, which indicates a strong 24-hour correlation. All in all, typical storage workloads follow a repeatable pattern and storage clusters are usually impacted by the number of time zones in which it is deployed. 

Given the strong correlations observed at hourly, daily and weekly lags, the IO arrival and block write rates emerge as time series models. The forecast produced from the time series model is used by the background scheduler to populate an hourly, daily, weekly, and even monthly plan of how system resources (i.e., CPU and memory resources) are allocated. This plan will guide the scheduler on how much background debt to defer (size the background debt buckets) and indicate the time they will be paid, that is based on forecasting the idle cycle, we can note time we start processing this debt.

The input sensors/features of our forecasting algorithm are:
\begin{itemize}
\item the IO arrival rate, provided at different granularities (different granularity may favor a different forecasting method).
\item the free storage capacity.
\item Types of load/bursts (percentage by load type: read, write, unmap).
\item Deferred background debt (averaged in block size) and per-queue processing rate.
\end{itemize}
The smart background scheduler predicts the idle cycles, level of idleness(remaining processing capacity), storage capacity, deferred debt, produce a debt processing plan, and ultimately size the deferred-debt bucket accordingly. The bucket size of the deferred-debt continuously grows and shrinks according to the debt processing plan. Growing the debt bucket allows the system to defer more background work and reduce the impact on foreground IOs during peak times. The scheduler will ultimately produce an hourly, daily, and weekly forecast plan of potential free windows when debt can be paid off. The window indicates how much debt can be paid and which debt to pay first. The scheduler prioritizes debt processing based on how much space can be reclaimed and ultimately guide a weekly debt balance sheet to keep debt in check.

\section{Methodology}
\subsection{Forecasting Model}
The IO arrival intensity exhibit many elements of a time series model. The incoming IO count and the read-write mix observed at any specified time window is a time ordered series. There is some great research already established in the area of time series forecast. We compare few of the leading methods that worked best in this area. However, we add new dimensions that are specific for our model such as forecasting read-write mix, free pool space, and the cost to clear Background debt. Background ops such as data/metadata flush requires a more finer time granularity (second/minute). \\
In particular, we explore and compare the following models which are sorted by compute-complexity. We expect the more compute-complex the model is, the better at prediction estimation and the more states it can identify:

\begin{itemize}
\item Time Series based on Markov Modulated Poisson Process \cite{Xue2014} (We identify user arrival intensity low/medium/high states and interval between them). This is no suitable for fine-grained  forecasting.
\item Time Series based on ARMA/ARIMA(p, d, q) model (Auto-Regression Integrated Moving-Average) \cite{chatfield2019analysis}. We calculate ACF and PACF to choose p \& q parameters which correspond to lag and moving-average window size, while d becomes a factor if trend is not stationary.
\item Time-Series based on Triple Exponential Smoothing using Holt-winters' method \cite{Goodwin} (Trend, Season, and Residual factors).
\item Machine Learning techniques, in particular, reinforcement learning (RL) and Long-Term-Short-Term-Memory (LSTM). This non-linear solution is to be explored for a finer prediction model. 
\end{itemize}

However, We propose a forecasting model that is similar to exponential smoothing. This method initially decomposes the time-series into trend, season, and noise where:\\
$Y_t = Trend_t + Season_t + Noise_t$ \\
Exponential smoothing forecasting model does not provide multiple seasonality periods. Trend index may have a weekly cyclic pattern but season index most likely have a daily cycle (see section 4). In order to predict future days’ IO intensity or read-write mix, we use an exponentially weighted moving average ($EWMA$) of the trend with a weekly cycle and season based on daily cycle where the $EWMA$ for a series $Y$ is defined as a recursive function:
\[
  S_t=\begin{cases}
               Y_1, \hspace{27mm} t=1\\
               \alpha Y_t + (1 - \alpha) S_{t-1}, \hspace{3mm} t>1
            \end{cases}
\]

Where:
\begin{itemize}
\item The coefficient $\alpha$ represents the degree of weighting decrease, a constant smoothing factor between 0 and 1. A higher $\alpha$ discounts older observations faster.
\item $Y_t$ is the value at a time period t.
\item $S_t$ is the value of the EWMA at any time period t.
\end{itemize}
The weekly forecast is simply the sum of trend index and season index:\\
$$\hat{Y_t} = EWMA_{Trend_t} + EWMA_{Season_t}$$
Even though the prediction error might still be high, MPE (Mean Percentage Error) is expected to be low. The white noise effect of over-prediction (positive error) and under-prediction (negative error) will offset each other out, where the prediction error at time $t$ is the following:

$$\sum_{0}^{t} (Y_t-\hat{Y_t}) = \epsilon$$

where $\epsilon$ is bounded by error at time $t$, i.e. ($Y_t-\hat{Y_t}$).\\
We are not concerned at minimizing the error between prediction and actual as long as the cumulative error at any time $t$ is small. This comes from the idea that if the prediction error lead to burning through debt more than required at any time t, the next cycle $(t+1)$ can lead to burning less debt which essentially cancel out the prediction error effect.

Predicting and factoring in the IO arrival rate, the type of IO (read, write, unmap, etc …),  the various data services policies, and the deferred debt is key to estimate free pool space and the potential freeable space if we execute an item off the background debt queue.

\subsection{Background Scheduler Model}
As we explained earlier, the scheduler needs the following data to make a good scheduling decision which become the features of the established learning algorithm:
\begin{itemize}
\item Forecast of IO arrival intensity (by type: read, write) and forecast of free pool space based on adopted data services policies.
\item Accumulated Debt of different background services, cost to service one item off each of the background queues, and potential tied capacity of an item off each of the BG queues.
\item Current free capacity and estimation of tied capacity.
\end{itemize}
The learning algorithm will produce the following information for the smart scheduler:
\begin{itemize}
\item Partition the resources between foreground and background processing
\item Prioritize the BG queues.
\item Provide a future plan of when and how much to process off the BG debt and size the various BG buckets accordingly.
\end{itemize}
\begin{figure}
  \includegraphics[width=0.5\textwidth]{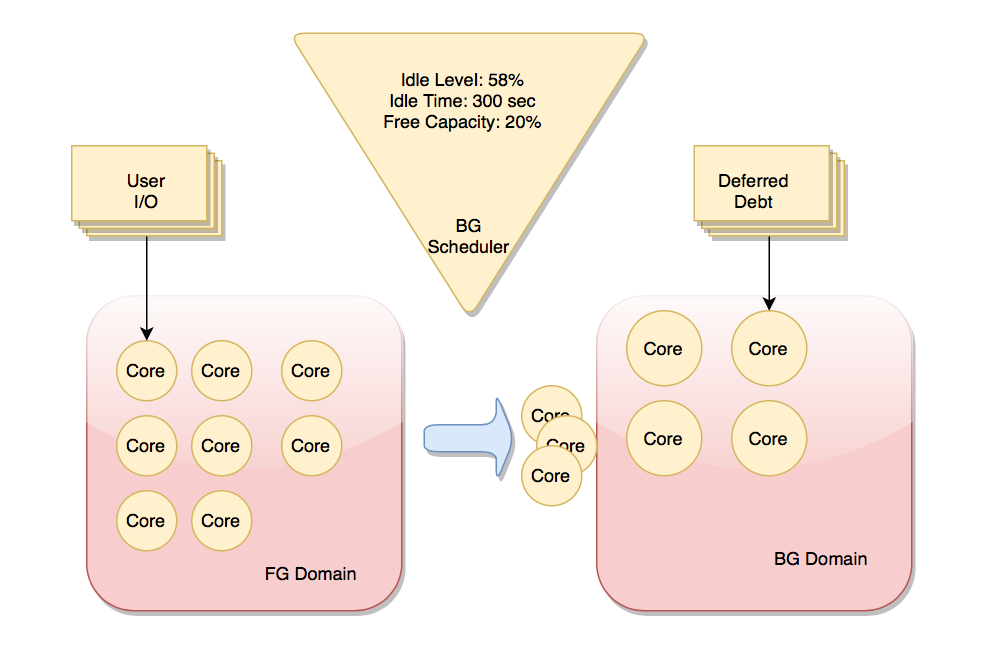}
  \caption{Background Scheduler executing isolated domains according to forecasted plan.}
  \label{fig:partitionmodel}
\end{figure}

Since the background scheduler may have to work in a muddied environment along with foreground I/O if the forecast is not favorable for the upcoming periods. This usually happen due to a stressed system condition. In order to avoid cache thrashing or varying runtime effects, we tend to execute background jobs in an isolated domain that either uses a separate CPU socket in multi-socket platforms or group of cores in a single socket platforms. Partitioning the resources and segregating foreground I/O from background ops allows foreground requests to execute in a lower latency environment and without the need to warm up the caches. This is a clear distinction and advantage of our priority based scheduler.  Figure \ref{fig:partitionmodel} shows a rough idea how to execute the isolated domains according to scheduler forecasted plan. Foreground I/O latency typically demands a relatively small latency compared to other ops. Having for foreground I/O to wait behind a heavy lifting and bulky background thread impacts foreground latency greatly. Thus, the importance of partitioning the resources and segregating cores used to service foreground I/O vs background is a key part of this design.

The foreground IO load is a combination of both read and write loads where read factor is usually multiple that of write.
$$L_{FG}(t) = L_{Read}(t) + L_{Write}(t)$$
$$L_{Read}(t) = r(t) * L_{FG}(t)$$
$$L_{Write}(t) = (1 – r(t)) * L_{FG}(t)$$
where $L_{FG}(t)$ is the foreground arrival IOPS and $r(t)$ is the read load ratio at any given time $t$.
Space capacity utilization is a cumulative function of previous time period.

$$U(t+1) = U(t) +  u(t) * L_{Write}(t) + D(t)$$
where $U(t+1)$ is the capacity utilization at future time $t+1$ and $u$ is the ratio of unique blocks, and $D(t)$ is deferred debt blocks at time $t$.\\
Each of the deferred debt types can be derived from foreground arrival rate using a unique formula\footnote{$BSize$ is the default block size, $Len$ is the length of unmap, and $DMD\_Ratio$ is the data to metadata ratio and is dependent on the namespace implementation of storage system and the efficiency ratio of its data-path. It varies but will consider to fix it between 2-5\% of total space utilization.} :

$$D_{Overwrite}(t) = (1 – u(t)) * L_{Write}(t)$$
$$D_{Unmap} (t) = \frac{Len (t)}{BSize} * (1 + DMD\_Ratio)$$
$$D_{Snap\_Delete}(t) = L_{Write}(t) * Snap\_Retention\_Time$$

Hardware can be modeled as a tuple of $<socket, core, memory>$. Each core is confined to either a foreground processing domain or background processing domain at any given time t. However, we will relax the memory element from the isolated domain for the following reasons:

\begin{itemize}
\item Most of today’s storage server memory is merely a big read cache. 
\item We assume that cache eviction rate is low due to background processing mainly working on metadata.
\item A good prefetch algorithm already deployed. 
\item Metadata is a tiny fraction (2-5\%) when compared to data. We assume that all metadata will fit in cache.
\end{itemize}

The forecaster is used to predict the foreground arrival intensity rate $L_{FG}(t)$ and unique block fraction $u(t)$ of foreground writes with a different granularity to cover both the short-term (minute/hour) and long-term (day/week) goals. These predictions are then used to drive the future debt $D(t)$ to be accumulated against required data services. 
For example, if policy is to create a scheduled snapshot every 1 hour for a group of 10 LUNs with retention time equivalent for 1 hour, then the debt accumulated is equivalent to: 
 
$$D_{Snap\_Delete} = 10 * \sum_{t}^{t+3600}L_{write}(t)$$
We assume, that $L_{Write}(t)$ targeting the 10 luns, is identical. The time to service one item of this debt queue is directly proportional to number of foreground writes:

$$RT_{Snap\_Delete} = c *D_{Snap\_Delete}$$
where $c$ is the cost to read and manipulate one block. 

Given that we modeled the cost to service the deferred debt, the capacity tied in the various debt queues, future forecast and future debt, we use multi-server queuing theory (M/M/c) to allocate the cores between foreground and background. However, the optimization is to suppress and segregate background processing and bias foreground work as long as capacity is not violated in the near future. The priority scheduler assigns priority on each of the background queues based on the expected reclaimed capacity of this queue, data service / reliability SLA (such as time to rebuild a drive), and time to service one item off this queue.

Each core can service some given throughput $CIOPS_{FG}$ (FG Core IOPS) at a preferable latency $RT_{FG}$. This number can be quickly obtained when evaluating a new storage server platform or on a live system by a brief calibration period where data services are paused. If the total system throughput demand per current arrival rate and the forecast arrival rate for the next $x$ seconds is $IOPS_{a}$, then:

$$C_{FG} (t) = MIN \{ N, \frac{IOPS_{a}}{CIOPS_{FG}} – CFF(t) \}$$
$$C_{BG} (t) = N - C_{FG} (t)$$
where $C_{FG} (t)$ and  $C_{BG} (t)$ is the number of cores allocated for foreground and background respectively for the next $x$ seconds. $N$ is the total cores available in system for Datapath. 

$CFF(t)$ is the capacity forecast factor. It is the extra cores we need to steal from the set servicing foreground IOs in order to guarantee pool never run of free space before the forecasted “long idle phase”. This $CFF(t)$ factor indeed impacts foreground IO and will ultimately lead to generating less background work. This is a sign of oversubscribed system where we record as a violation. If said violations increase dramatically over time, the operator is notified to add a new disk to pool or reduce/balance application load.

Since we have $n$ number of background queues with a tuple of $< servicing time, priority >$ and $m$ number of background cores (servers), we can use the multi-server queuing theory $(M/M/c)$ to quantify if we need to add more servers. We can limit the bucket size of each of the queues based on its priority (tied capacity) and limit total bucket size of all the queues based on forecasted free capacity. The scheduler round robin between the various background queue to pull $x$ items of each queue based on its priority and push it to each core.

\section{Results and Discussion}
\subsection{Workload characterization and forecasting}
\begin{figure}[h!]
  \includegraphics[width=0.5\textwidth]{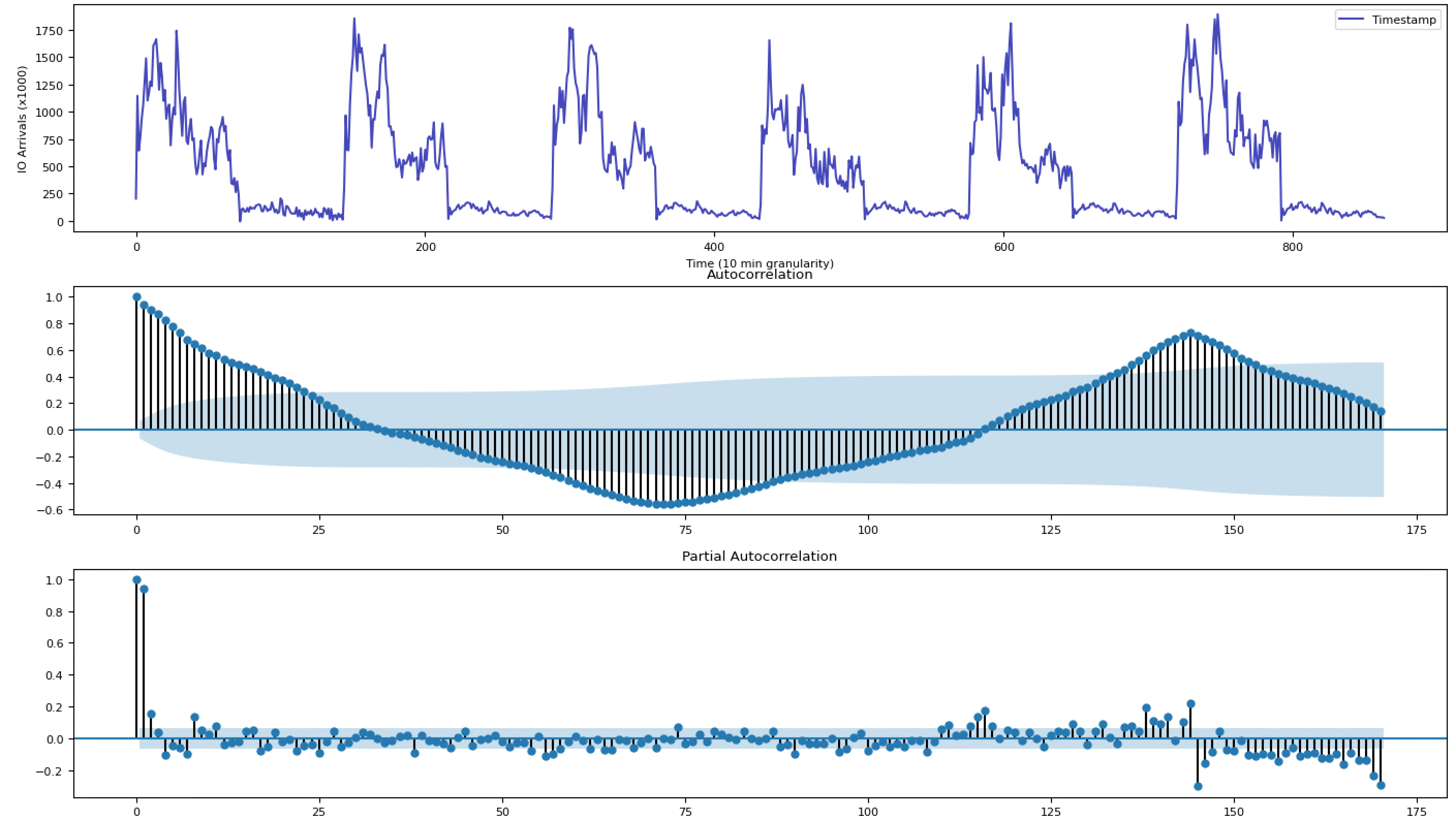}
  \caption{ACF and PACF of VDI Workload IO Arrival Intensity.}
  \label{fig:acf-pcf}
\end{figure}

\begin{figure}[h!]
  \includegraphics[width=0.5\textwidth]{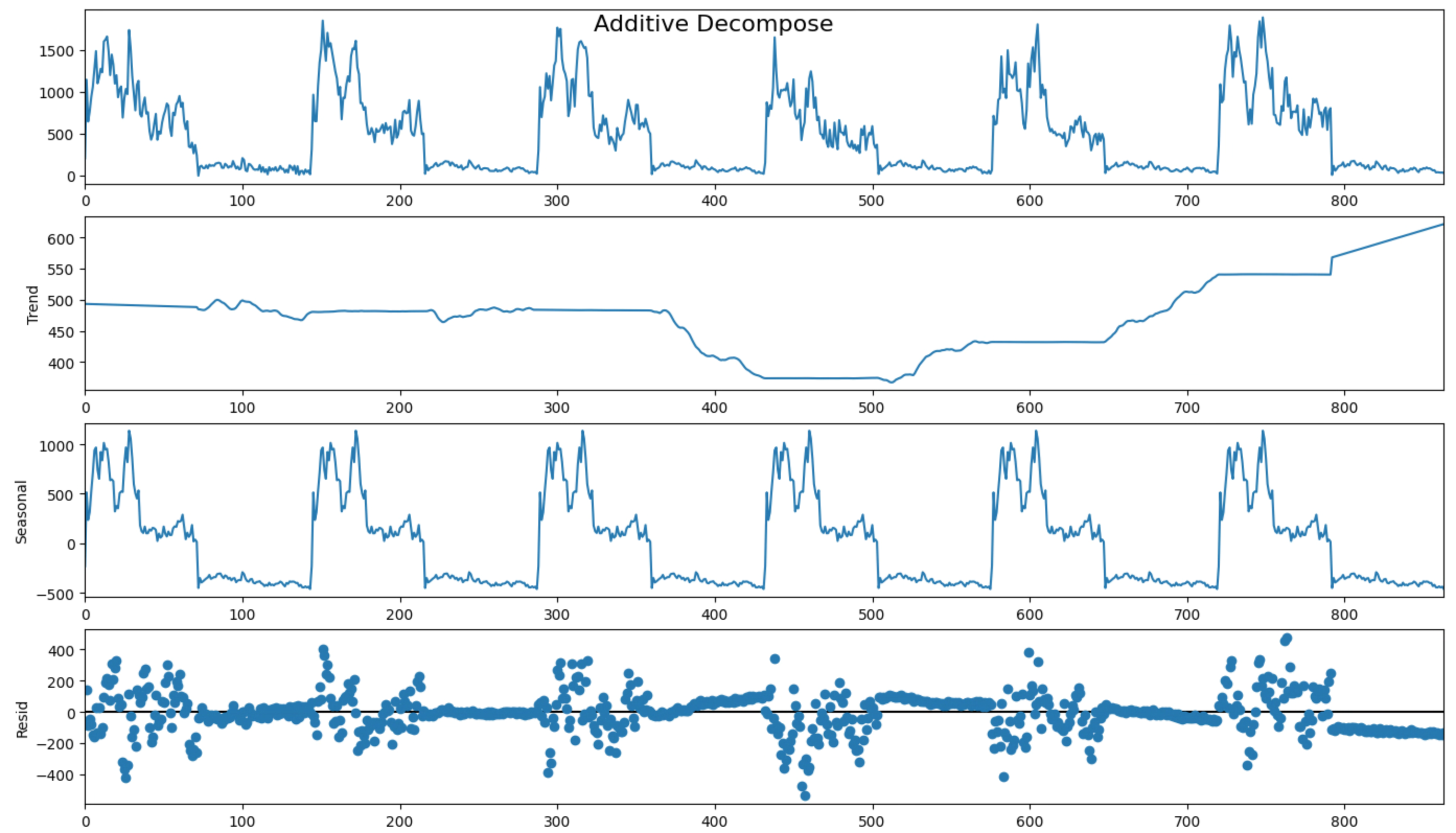}
  \caption{Additive Decompose (trend, season, and residual) of VDI IO Arrival Intensity.}
  \label{fig:decompose}
\end{figure}

Figure \ref{fig:acf-pcf} shows the ACF and PACF of VDI workload IO Arrival Intensity. It shows strong correlation at 24-hour lag. However, the intensity signal is not always stationary due to multiple seasonality effect such as weekend and holidays. In order to reach a stationary intensity signal, we group weekday as a separate cluster from weekend, leading to 2 stationary intensity signals one for weekday and another one for weekend. However, the signal additive decompose of VDI workload as shown in Figure \ref{fig:decompose} shows a stationary trend for Wednesday through Friday, a lower intensity for weekend, and higher intensity on Monday. Thus, we suggests to utilize a stationary trend factor for each cluster of days using a K-means clustering method with upper bound of \( \sqrt{\frac{n}{2}} \) where n is the number of days in the week in addition to holidays.


\begin{figure}[h!]
  \includegraphics[width=0.5\textwidth]{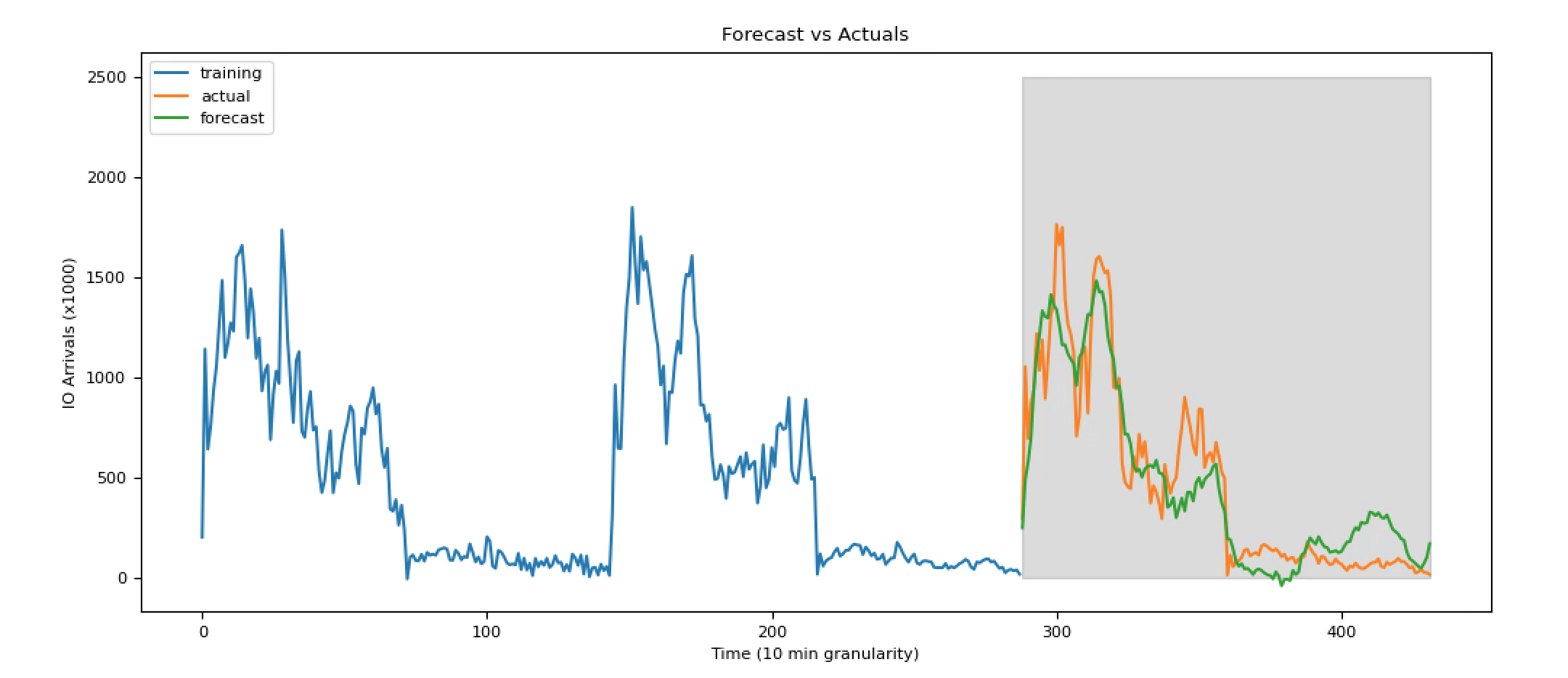}
  \caption{ARIMA forecasting performance on a cluster of 3 days.}
  \label{fig:arima}
\end{figure}

Initially, we used the standard ARIMA model to train on a cluster of days that has a stationary trend property, where the autoregressive factor (p) set to 144 (24 hours @ 10-minute granularity), the differential factor (d) set to 0,  and the moving average factor (q) set to 2. Figure \ref{fig:arima} shows forecast vs actual of IO arrival intensity and write intensity using ARIMA model. ARIMA achieves Symmetric mean absolute percentage error (SMAPE) of 27.8\% when comparing prediction to actual.

\begin{figure}[h!]
  \includegraphics[width=0.5\textwidth]{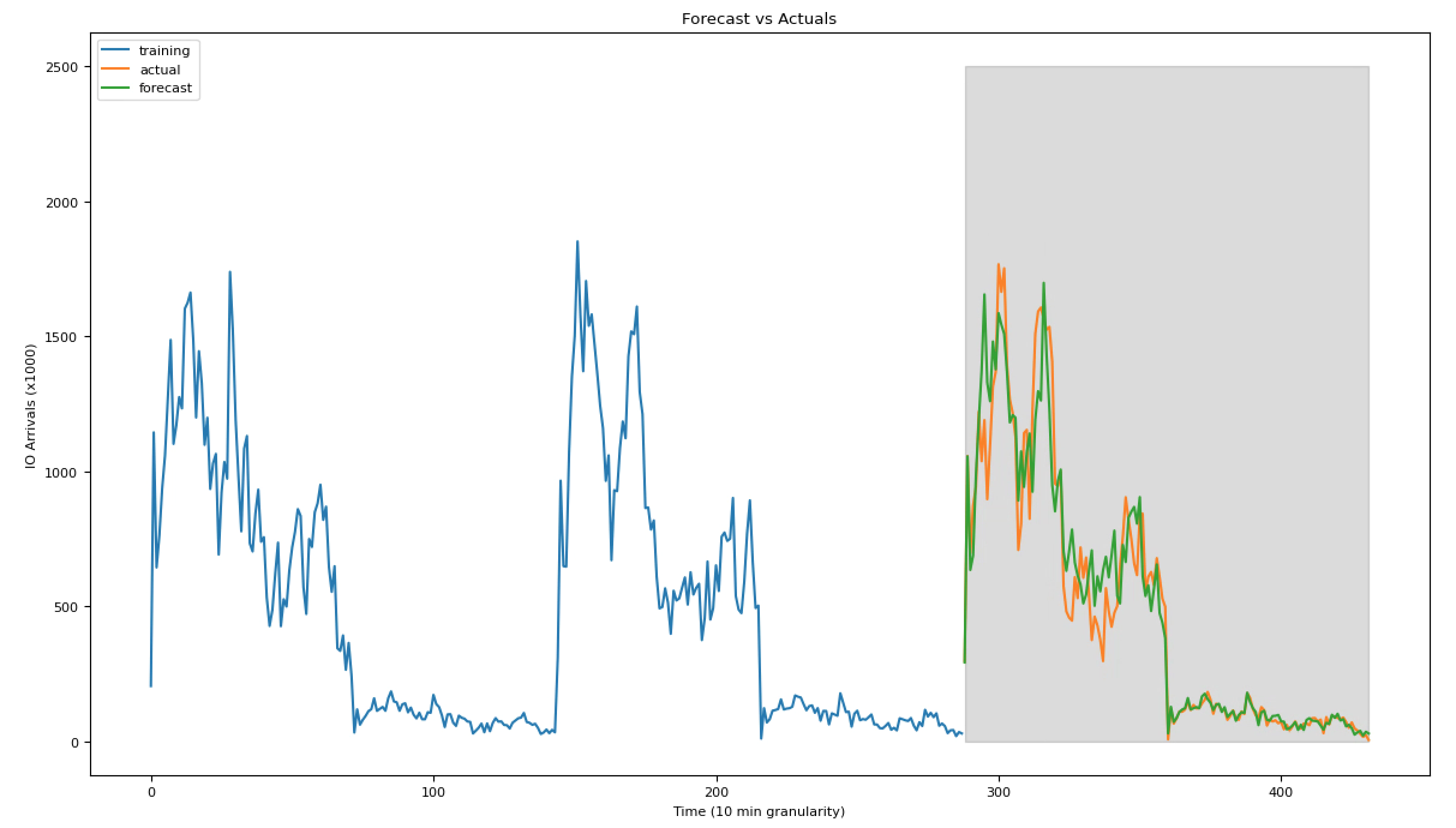}
  \caption{Prediction by Triple Exponential Smoothing.}
  \label{fig:smoothing}
\end{figure}

We then experimented with a triple exponential smoothing model where season period is set to 144 (24 hour @ 10 minute granularity) similar to ARIMA and a damped trend option. Figure \ref{fig:smoothing} shows forecast vs actual of IO arrival intensity using triple exponential smoothing. SMAPE achieved for triple exponential smoothing is 19.2\%.

\begin{figure}[h!]
  \includegraphics[width=0.5\textwidth]{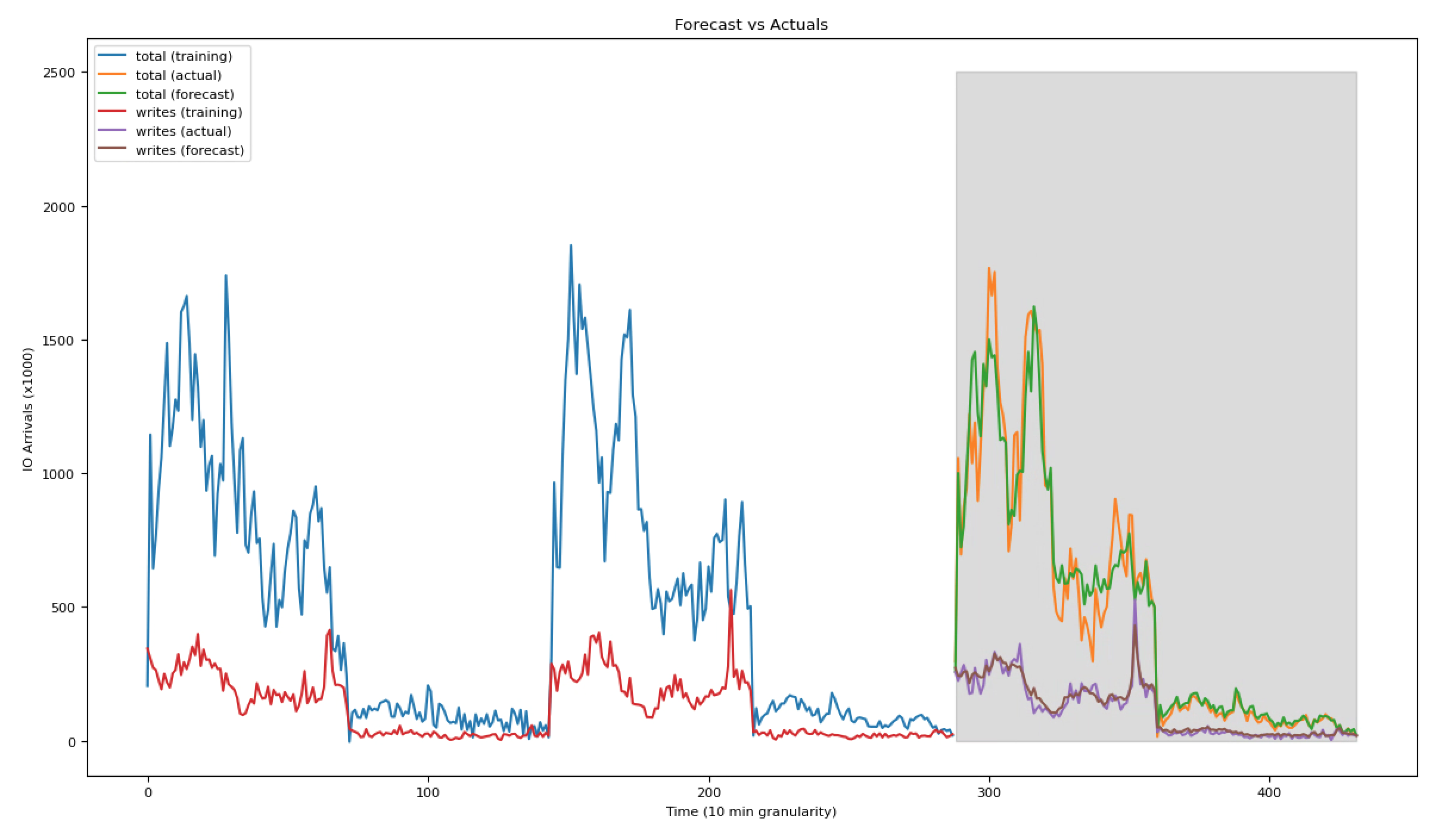}
  \caption{Prediction by averaging trend and season of historical week's data.}
  \label{fig:2signal}
\end{figure}

Finally, we experimented with our simple but robust forecasting model. Figure \ref{fig:2signal} shows forecast vs actual of IO arrival intensity and write intensity using our method. We achieve Symmetric mean absolute percentage error (SMAPE) of 14.8\% prediction to actual which means we cut the error range by half if compared to ARIMA and few percentage better than triple exponential smoothing.

Even though the prediction error is still over 10\%, MPE (Mean Percentage Error) is well below 10\% as the white noise effect of over-prediction (positive error) and under-prediction (negative error) offsets each other out as discussed in a previous section.

\subsection{Smart Background Scheduler}

\begin{figure}
  \includegraphics[width=0.5\textwidth]{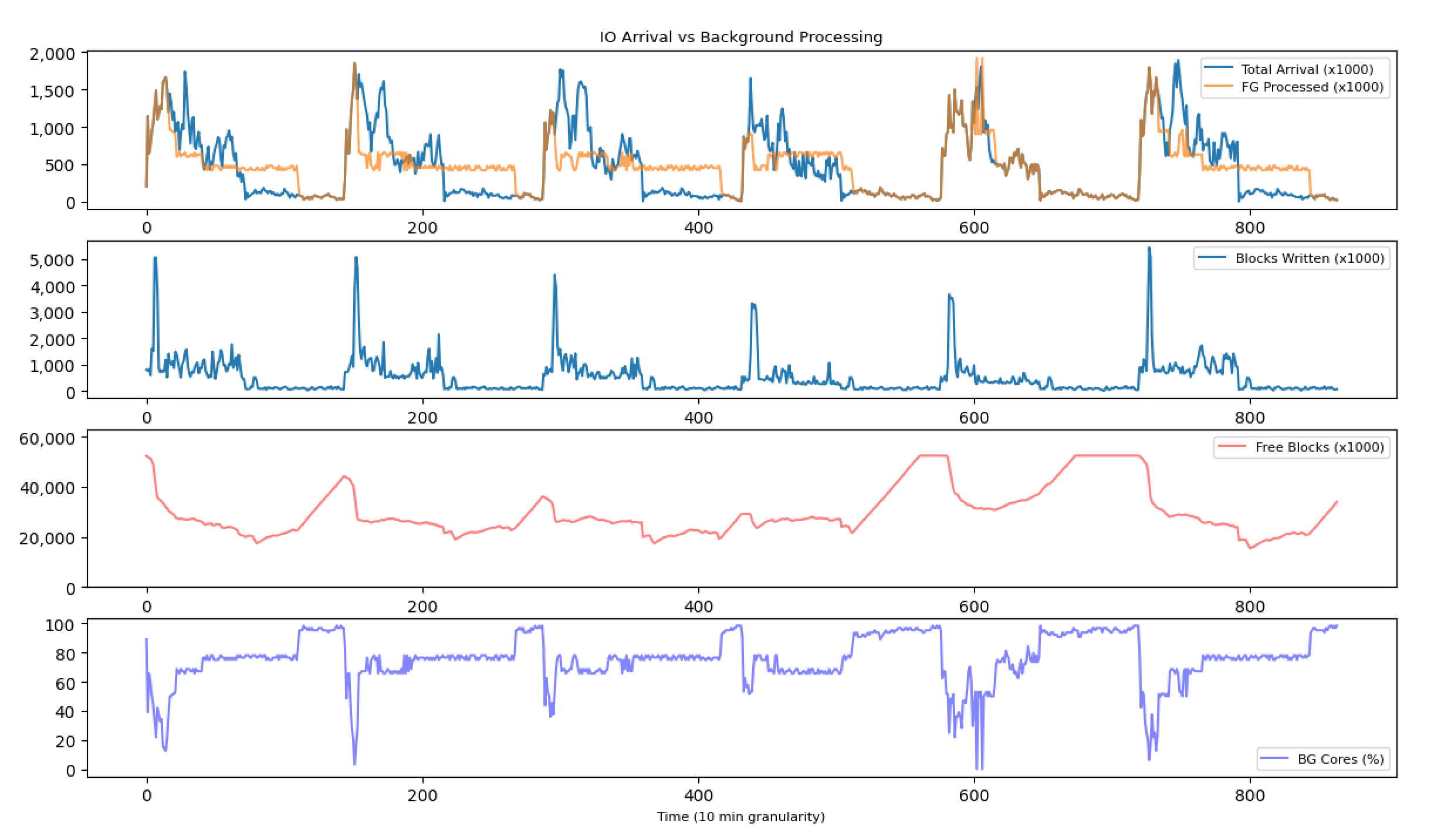}
  \caption{Performance of BG Scheduler when using fixed debt bucket size.}
  \label{fig:baseline}
\end{figure}

We compared 2 implementations of the background scheduler. Both implementations include replaying the VDI traces of 3 LUNs for the 6-day period with a snap creation/retention time of 1 hour. We fix the number of cores in simulation to 64, the per core FG processing rate to 50 IOPS, BG processing rate to 20 ops per second, and a 4K block size. The first implementation used a fixed debt bucket size. We fix the debt bucket size to a range between 40 and 50\% of total pool size where the scheduler is less aggressive at burning the debt at 40\% and become more aggressive as we reach 50\%. Such an implementation leads to SLO performance violation of 54.6\% where 21.7\% of IO was queued due to out of resources as shown in Figure \ref{fig:baseline}.

\begin{figure}
  \includegraphics[width=0.5\textwidth]{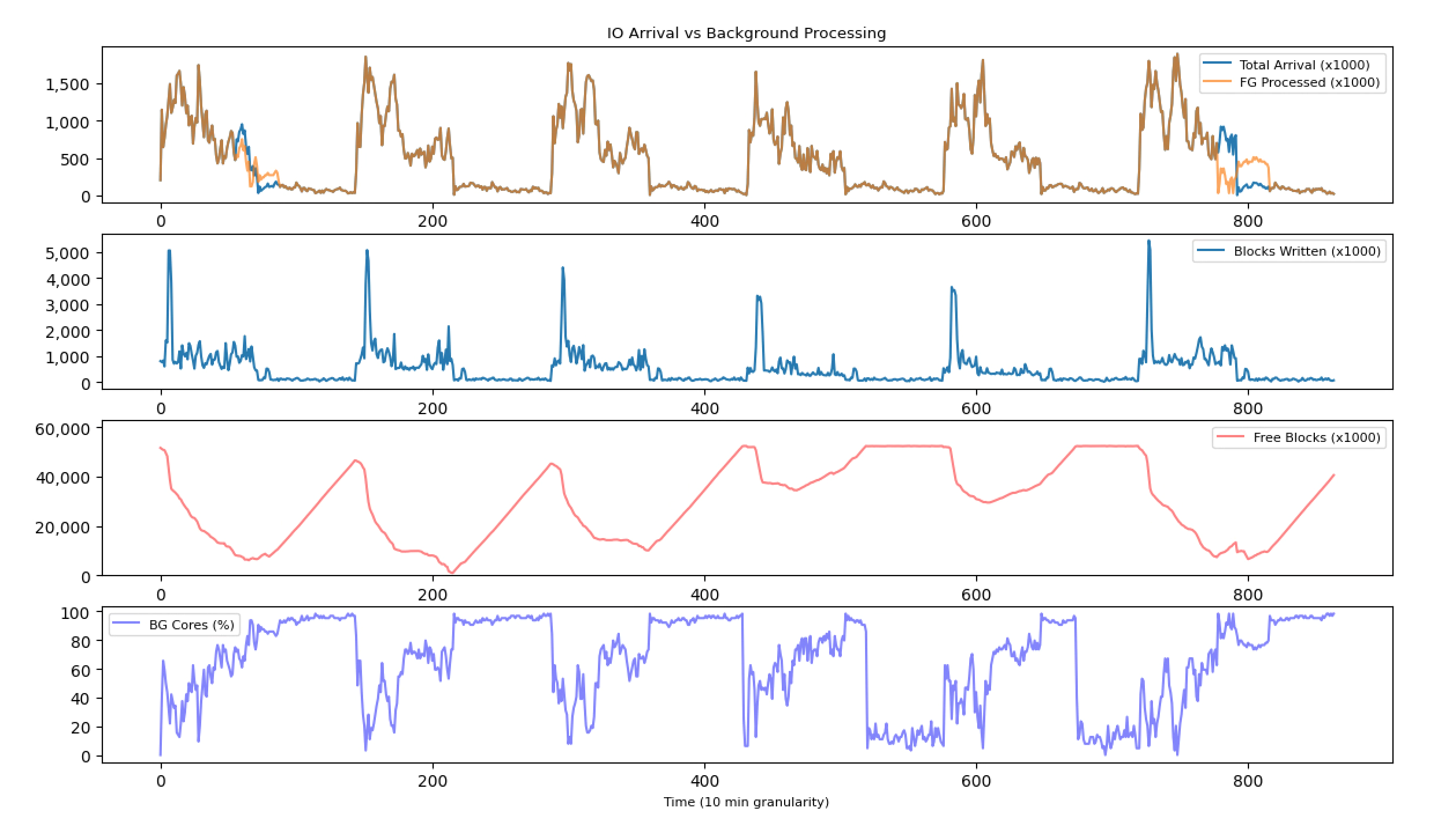}
  \caption{Performance of BG Scheduler utilizing forecasted information.}
  \label{fig:smart}
\end{figure}

The second implementation uses a dynamic bucket size that utilizes the forecasted information described in section 5. The debt bucket implementation can grow up to 95\% of pool size as long as the forecasted data allows. The smart scheduler algorithm utilizes the forecasted data for the upcoming period (example: week) to replay the forecasted foreground IO, background IO, free pool capacity, and debt size. The dynamic bucket algorithm maximizes foreground I/O as long there are free pool capacity. If a future period is observed where pool capacity will be 95\% depleted, a greedy algorithm evenly spreads enough background load to a relative prior period. This is necessary in order not to run out of pool space in the future as we maximize foreground IO. Several iterations are needed to replay the forecasted data series and slowly increase the background processing  rate so to find the optimal partition point between foreground and background cores. With this scheme we are able to achieve SLO performance violation of only 6.2\% through the 6 days where 2.6\% of IO was queued due to out of resources as shown in Figure \ref{fig:smart}.

\section{Conclusion and Future Work}
In this paper, we have proposed a smart background scheduler for storage systems that harvest every ounce of processing capacity to absorb application bursts and meet SLOs. The design requires 2 modular pieces to work hand in hand to get this done. The first module provides the forecast of information such as IO arrival intensity, data blocks written, and unique data blocks written. The second module uses the forecasted data for the upcoming days to partition the resources between foreground and background ops biasing foreground IO as long as pool capacity is not depleted or reliability SLA is not violated. Through replaying VDI workload traces, the smart background scheduler reduced SLO violations from 54.6\% when using a fixed debt bucket to merely a 6.2\%.

These items are still in progress and currently constitutes the future work:

\begin{enumerate}
    \item Compare the proposed forecasting method to recurrent neural network method such as LSTM and ensembles of time series methods.
    \item Background ops such as data/metadata flush requires a more finer time granularity (second/minute) and our forecasting/scheduling method might need tweaking to address this kind of environment. Compare to methods from literature.
    \item Implement other background ops such as efficiency ops (offline deduplication and compression) or reliability ops (drive rebuild).
    \item Provide and demonstrate the priority based scheduler on multiple op types.
\end{enumerate}

\bibliographystyle{IEEEtran}
\bibliography{Scheduler}

\begin{thebibliography}{10}
\providecommand{\url}[1]{#1}
\csname url@samestyle\endcsname
\providecommand{\newblock}{\relax}
\providecommand{\bibinfo}[2]{#2}
\providecommand{\BIBentrySTDinterwordspacing}{\spaceskip=0pt\relax}
\providecommand{\BIBentryALTinterwordstretchfactor}{4}
\providecommand{\BIBentryALTinterwordspacing}{\spaceskip=\fontdimen2\font plus
\BIBentryALTinterwordstretchfactor\fontdimen3\font minus
  \fontdimen4\font\relax}
\providecommand{\BIBforeignlanguage}[2]{{%
\expandafter\ifx\csname l@#1\endcsname\relax
\typeout{** WARNING: IEEEtran.bst: No hyphenation pattern has been}%
\typeout{** loaded for the language `#1'. Using the pattern for}%
\typeout{** the default language instead.}%
\else
\language=\csname l@#1\endcsname
\fi
#2}}
\providecommand{\BIBdecl}{\relax}
\BIBdecl

\bibitem{mit}
\BIBentryALTinterwordspacing
{MIT Genius Stuffs 100 Processors Into Single Chip | WIRED}. [Online].
  Available: \url{https://www.wired.com/2012/01/mit-genius-stu/}
\BIBentrySTDinterwordspacing

\bibitem{dell}
\BIBentryALTinterwordspacing
{Dell EMC PowerStore Released For Midrange - StorageReview.com}. [Online].
  Available:
  \url{https://www.storagereview.com/news/dell-emc-powerstore-released-for-midrange}
\BIBentrySTDinterwordspacing

\bibitem{Qiao2019}
Z.~Qiao, S.~Fu, H.~B. Chen, and B.~Settlemyer, ``{Building Reliable
  High-Performance Storage Systems: An Empirical and Analytical Study},''
  \emph{Proceedings - IEEE International Conference on Cluster Computing,
  ICCC}, vol. 2019-September, 2019.

\bibitem{wiki01}
\BIBentryALTinterwordspacing
{List of Log Structured File Systems}. [Online]. Available:
  \url{https://en.wikipedia.org/wiki/List\_of\_log-structured\_file\_systems}
\BIBentrySTDinterwordspacing

\bibitem{Xue2014}
J.~Xue, F.~Yan, A.~Riska, and E.~Smirni, ``{Storage Workload Isolation via Tier
  Warming: How Models Can Help},'' \emph{11th International Conference on
  Autonomic Computing (ICAC 14)}, pp. 1--11, 2014.

\bibitem{Riska2006}
A.~Riska and E.~Riedel, ``{Long-range dependence at the disk drive level},''
  \emph{Third International Conference on the Quantitative Evaluation of
  Systems, QEST 2006}, no. July, pp. 41--50, 2006.

\bibitem{Yan2012}
F.~Yan, A.~Riska, and E.~Smirni, ``{Busy bee: How to use traffic information
  for better scheduling of background tasks},'' \emph{ICPE'12 - Proceedings of
  the 3rd Joint WOSP/SIPEW International Conference on Performance
  Engineering}, pp. 145--156, 2012.

\bibitem{Eggert2005}
L.~Eggert and J.~D. Touch, ``{Idletime scheduling with preemption intervals},''
  \emph{Proceedings of the 20th ACM Symposium on Operating Systems Principles,
  SOSP 2005}, pp. 249--262, 2005.

\bibitem{Stokely2012}
M.~Stokely, A.~Mehrabian, C.~Albrecht, F.~Labelle, and A.~Merchant,
  ``{Projecting disk usage based on historical trends in a cloud
  environment},'' \emph{ScienceCloud '12 - 3rd Workshop on Scientific Cloud
  Computing}, pp. 63--70, 2012.

\bibitem{Mi2012}
N.~Mi, G.~Casale, and E.~Smirni, ``{ASIdE: Using autocorrelation-based size
  estimation for scheduling bursty workloads},'' \emph{IEEE Transactions on
  Network and Service Management}, vol.~9, no.~2, pp. 198--212, 2012.

\bibitem{Taylor2017}
S.~J. Taylor and B.~Letham, ``{Business Time Series Forecasting at Scale},''
  \emph{PeerJ Preprints 5:e3190v2}, vol.~35, no.~8, pp. 48--90, 2017.

\bibitem{Alshawabkeh2012}
M.~Alshawabkeh, A.~Riska, A.~Sahin, and M.~Awwad, ``{Automated storage tiering
  using markov chain correlation based clustering},'' \emph{Proceedings - 2012
  11th International Conference on Machine Learning and Applications, ICMLA
  2012}, vol.~1, pp. 392--397, 2012.

\bibitem{Ravandi2017}
B.~Ravandi, I.~Papapanagiotou, and B.~Yang, ``{A Black-Box Self-Learning
  Scheduler for Cloud Block Storage Systems},'' \emph{2016 IEEE 9th
  International Conference on Cloud Computing (CLOUD)}, pp. 820--825, 2017.

\bibitem{Ravandi2017a}
B.~Ravandi and I.~Papapanagiotou, ``{A Self-Learning Scheduling in Cloud
  Software Defined Block Storage},'' \emph{IEEE International Conference on
  Cloud Computing, CLOUD}, vol. 2017-June, pp. 415--422, 2017.

\bibitem{Xue2016}
J.~Xue, F.~Yan, A.~Riska, and E.~Smirni, ``{Scheduling data analytics work with
  performance guarantees: queuing and machine learning models in synergy},''
  \emph{Cluster Computing}, 2016.

\bibitem{Zhang2006}
Q.~Zhang, A.~Riska, N.~Mi, E.~Riedel, and E.~Smirni, ``{Evaluating the
  performability of systems with background jobs},'' \emph{Proceedings of the
  International Conference on Dependable Systems and Networks}, vol. 2006, pp.
  495--504, 2006.

\bibitem{fio}
\BIBentryALTinterwordspacing
{FIO Benchmark}. [Online]. Available:
  \url{http://www.freecode.com/projects/fio}
\BIBentrySTDinterwordspacing

\bibitem{snia}
\BIBentryALTinterwordspacing
{SNIA traces}. [Online]. Available: \url{http://iotta.snia.org/tracetypes/3}
\BIBentrySTDinterwordspacing

\bibitem{Urdaneta2009}
G.~Urdaneta, G.~Pierre, and M.~van Steen, ``{Wikipedia workload analysis for
  decentralized hosting},'' \emph{Computer Networks}, 2009.

\bibitem{chatfield2019analysis}
C.~Chatfield and H.~Xing, \emph{The analysis of time series: an introduction
  with R}.\hskip 1em plus 0.5em minus 0.4em\relax CRC press, 2019.

\bibitem{Goodwin}
\BIBentryALTinterwordspacing
P.~Goodwin, ``The holt-winters approach to exponential smoothing: 50 years old
  and going strong,'' \emph{Foresight: The International Journal of Applied
  Forecasting}, no.~19, pp. 30--33, 2010. [Online]. Available:
  \url{https://EconPapers.repec.org/RePEc:for:ijafaa:y:2010:i:19:p:30-33}
\BIBentrySTDinterwordspacing

\end{thebibliography}

\end{document}